\documentclass[twocolumn,showpacs,preprintnumbers,amsmath,amssymb]{revtex4}

\usepackage{graphicx}
\usepackage{dcolumn}
\usepackage{bm}
\usepackage{booktabs}
\usepackage{amsmath}
\usepackage{color}
\usepackage[]{hyperref}
\hypersetup{linkcolor=blue,urlcolor=cyan,
           anchorcolor=true,citecolor=true}
 
\newcommand{\U}[1]{\ensuremath{\mathrm{#1}}}

\newcommand{\fig}[1]{Fig.~\ref{#1}}

\renewcommand{\vec}[1]{\mathbf{ #1 }}

\renewcommand{\vec}[1]{\mathbf{ #1 }}

\newcommand{\psiop}[1]{{\psi}(\mathbf{#1})}

\renewcommand{\t}[1]{{#1}}
\renewcommand{\tt}[1]{{#1}}
\newcommand{\tc}[1]{{#1}}
\newcommand{\tcc}[1]{{#1}}
\newcommand{\tccc}[1]{{#1}}
\begin{document}


\title{$U(1)$-Symmetry breaking and \t{violation} of axial symmetry in $\U{TlCuCl_3}$ \tc{and other insulating spin systems}}

\author{Raffaele Dell'Amore}
\author{Andreas Schilling}%
\affiliation{%
 Physik-Institut, University of Zurich, Winterthurerstrasse 190, 8057 Zurich, Switzerland
}%
\author{Karl Kr\"amer}%
\affiliation{%
Department of Chemistry and Biochemistry, University of Bern, 3000 Bern 9, Switzerland
}%

\date{\today}
\begin{abstract}
We describe the  Bose-Einstein condensate of magnetic bosonic quasiparticles in insulating spin systems using a phenomenological  standard functional method for $T$ = 0. We show that results that are already known from advanced computational techniques immediately follow. The inclusion of a perturbative anisotropy term that violates the axial symmetry allows us to remarkably well explain a number of experimental features of the dimerized spin-1/2 system $\mathrm{TlCuCl_3}$. Based on an energetic argument we predict a general intrinsic instability of an axially symmetric magnetic condensate towards a violation of this symmetry, which leads to the spontaneous formation of an anisotropy gap in the energy spectrum \t{above the critical field}. \tc{We, therefore, expect that a true Goldstone mode in insulating spin systems, i.e., a strictly linear energy-dispersion relation down to arbitrarily small excitations energies, cannot be observed in any real material.} 
\end{abstract}

\pacs{75.10.Hk, 75.45.+j, 75.30.Gw, 73.43.Nq}
\maketitle

\section{\label{Intro}Introduction\protect\\ }
The concept of Bose-Einstein condensation (BEC), i.e., the occupation of a single quantum state by a macroscopic number of bosons, has been extended from real bosonic particles to
various types of quasiparticles with integer total spin \cite{Affl, giacit, Nik, rueg, wachter, mats, polar, magnon, snoke1, snoke2, magnon2}. Such discrete magnetic, electronic or lattice excitations are then treated as a set of bosons forming a Bose gas \cite{snoke1}. These quasiparticles usually possess a small effective mass, which permits to study BEC even at room temperature \cite{magnon2}. Both experiment \cite{harada, snoke2} and theory  \cite{haugst, andersen} suggest that the occurrence of a BEC in a 3D interacting Bose system has its origin in the spontaneous breaking of the $U(1)$ symmetry, thereby leading to a gapless and linear excitation spectrum in
the long-wavelength limit, i.e., to a Goldstone mode \cite{goldstone}. 
\\ Experimental observations in a number of quantum spin systems  can be explained within the theory of BEC, e.g., by a condensation of triplet states in dimerized spin-1/2 systems (hereafter called triplons) such as $\U{TlCuCl_3}$ \cite{Nik, rueg, giaandrueg}. The bosonic character of the these magnetic quasiparticles  allows to describe this spin-dimer system as a weakly interacting Bose gas. Inelastic neutron-scattering measurements in the condensate phase of $\U{TlCuCl_3}$  revealed, in accordance with theoretical investigations \cite{matsumoto}, the presence of a seemingly gapless and linear excitation spectrum down to very low excitation energies of the order of 0.75 meV in $\mu_0H$~=~14~T \cite{rueg}. This observation has been interpreted as a manifestation of \t{the} Goldstone mode.
\\ In the last years various scenarios for the consequences of anisotropy on the properties of a
magnetic BEC have been discussed in detail \cite{kolezhuk, glazkov, sirker2, sirker, sebastian}. The presence of any kind of anisotropy will, in principle, explicitly break the rotational (i.e, axial) symmetry of the bosonic system \cite {sirker}. The degree of spontaneous $U(1)$-symmetry breaking then depends on the order of magnitude of the anisotropic terms compared to the energy scale associated with the isotropic interactions
of the Bose gas \cite{sebastian}. The relatively large triplon bandwith in $\mathrm{TlCuCl_3}$ \cite{cava} exceeds the spin-nonconserving terms, such as an intradimer exchange (IE) anisotropy and a Dzyaloshinsky-Moriya (DM) anisotropy, by more than two orders of magnitude  \cite{kolezhuk, glazkov, sirker}. Nonetheless, the question to what extent the existing anisotropies in $\mathrm{TlCuCl_3}$ do affect the magnetic phase diagram, the Goldstone mode and other measurable quantities of $\mathrm{TlCuCl_3}$ is still an issue under investigation \cite{glazkov, kolezhuk, vyaselev, sirker}.
 \\By taking a perturbative anisotropy term into account we will consider in the following the influence of an IE-like anisotropy that \tt{explicitly} violates the axial symmetry, and we will study the consequences on the condensate phase of $\mathrm{TlCuCl_3}$. The influence of a possible DM-like anisotropy \cite{sirker, sirker2} is not considered here \cite{DM}. \t{Based on an energy consideration we will then argue that, as a consequence of an unavoidable magnetoelastic coupling,} even an axially symmetric magnetic system is unstable towards a spontaneous violation of this symmetry as soon as the BEC-state is formed.
 
\section{\label{susc}Functional method}
We describe the condensate at $T$~ =~ 0 with a macroscopic wave function, a complex scalar field $\psiop{\vec{r},t}$. \tt{Standard} \t{functional} methods used to describe a dilute Bose gas in the classical limit at $T$ = 0 yield an extremal condition for the potential energy \t{per dimer} \cite{andersen}, namely 
\begin{equation}
u(\psi) = -\mu\psi^\dag\psi + \frac{v_0}{2}\left(\psi^\dag\psi\right)^2 _{\vert_{\psi=\psi_0}}= min.,
\end{equation}
where $\mu$ is the chemical potential, $v_0$ is a constant related to a repulsive short-range
interaction \cite{haugst}, and $\psi^\dag$ is the complex conjugate of $\psi$. The minimum value $\psi_0$ then determines the condensate \t{fraction} $n_c(0) = \psi_0^\dag\psi_0$, \t{here defined as $n_c(0) = N_c/N_d$ with $N_c$ the number of condensed triplons and $N_d$ the number of dimers.}
 \\ In a dimerized antiferromagnet, the chemical potential is $\mu=g\mu_B\mu_0(H-H_c),$ where $\mu_B$ is the Bohr magneton, $g$ the Land\'e $g$-factor, and $H_c$ the critical magnetic field beyond which a triplet $S =1$ state is energetically equally favorable as the singlet $S = 0$ state. This can be expressed in terms of the energy gap $\mathit{\Delta}$ = $g\mu_B\mu_0H_c$ separating at zero field the $S = 1$ and the $S = 0$ states, respectively. 
\\In the case of an explicitly violated axial symmetry we may include a perturbation term $\vert \tilde{\gamma} \vert \left(\psi\psi + \psi^\dag\psi^\dag \right)$ \cite{sirker, Affl} to the potential energy that can arise in a real magnetic system from various sources such as anisotropic intra- and interdimer interaction constants $J$ and $\tilde{J}$, respectively. For  $\mathrm{TlCuCl_3}$ we have, for example, $\mu_0H_c \approx$ 5.6 T \cite{shira}, $v_0/k_B$ = 315 $\mathrm{K}$ \cite{yamada}, and a $\vert \tilde{\gamma} \vert$ of the order of 0.01 meV  \cite{kolezhuk, sirker}, depending on the orientation of the magnetic field $\vec{H}$ with respect to the crystal lattice \cite{kolezhuk}. Such an anisotropy term may arise from a pre-existing \t{violated} axial symmetry of the system, or from a spontaneous distortion at the magnetic phase transition that we will discuss below.
We, therefore, have to minimize
\begin{equation}
u(\psi) = -\mu\psi^\dag\psi + \vert \tilde{\gamma} \vert \left(\psi\psi + \psi^\dag\psi^\dag \right) + \frac{v_0}{2}\left(\psi^\dag\psi\right)^2,
\end{equation}
where we assume that $H_c$ itself is at first unchanged by the presence of the small perturbative anisotropy $\vert \tilde{\gamma}\vert \ll$~$\mathit{\Delta}$. 
\\We first want to compare the results of \t{this} simple \t{formalism} with corresponding predictions from advanced Hartree-Fock (HF) computations and with experimental data on $\mathrm{TlCuCl_3}$. Despite the simple formalism used here, we can later make specific predictions that would otherwise be more difficult to obtain.\\
\section{\label{Intro}Results \protect\\ }
\subsection{\label{Intro}Comparison with results from Hartree-Fock calculations \protect\\ }
\begin{figure}
\includegraphics[scale=0.36]{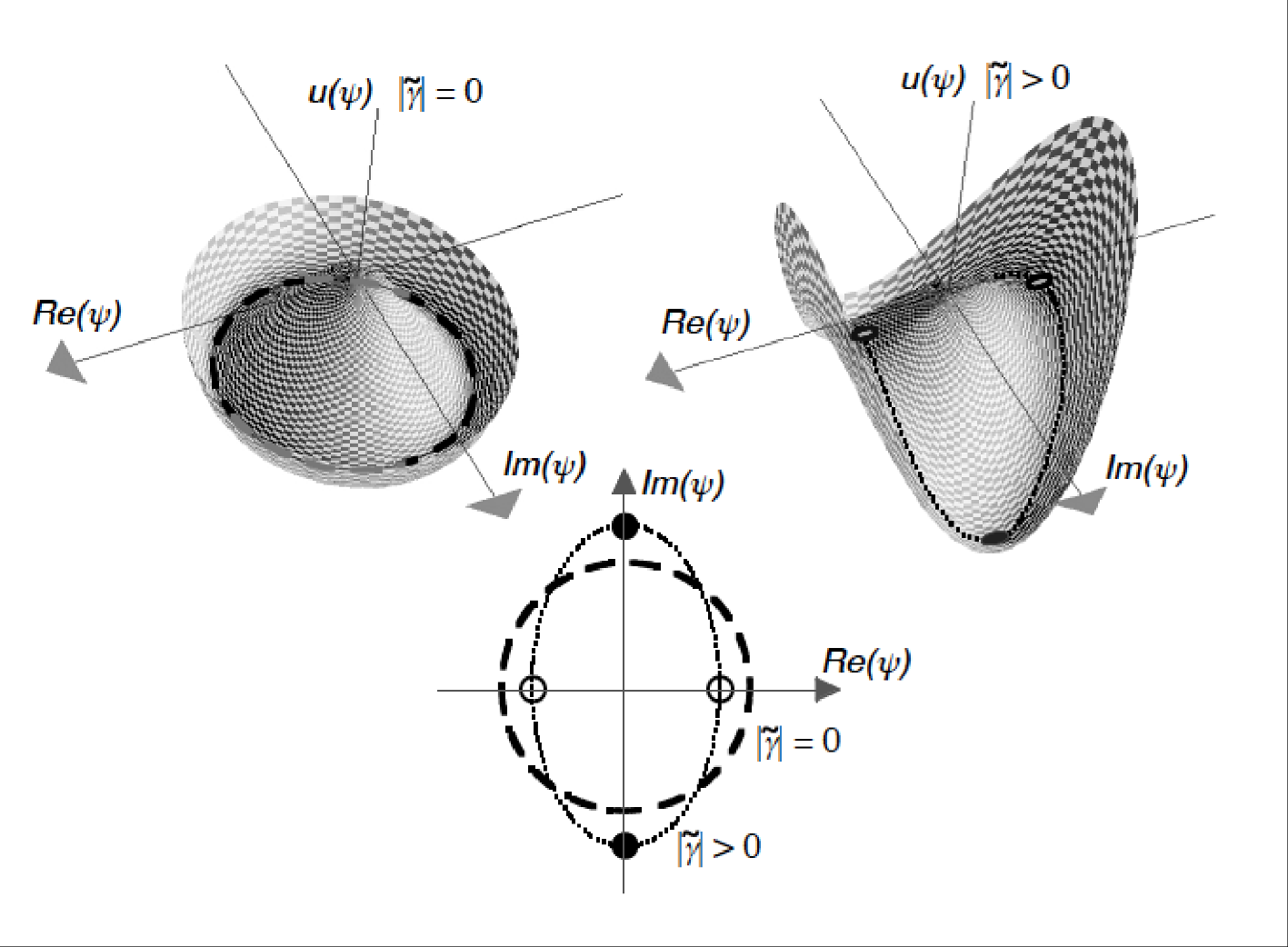}
\caption{\label{ls} Potential-energy $u$ as a function of $\psi$ for axial
symmetry (left panel, "mexican-hat potential") and violated axial symmetry (right panel, "Napoleon's hat potential"), respectively. In the symmetric case the minimum value $u_{min}$ = $-\mu^2/2v_0$ is realized along a circle (dashed line), while in the anisotropic case isolated minima $u_{min} = -(\mu+2\vert \tilde{\gamma}\vert)^2/2v_0$ are on the imaginary axis (filled circles).}
\end{figure}

Without any explicit anisotropy $\left(i.e., \vert \tilde{\gamma} \vert = 0\right)$ we obtain the well-known minimum value $\psi^\dag_0\psi^{}_0$ = $n_c(0$) = $\mu/v_0$ \cite{Nik}. The phase $\phi$ of $\psi_0=\vert \psi_0 \vert e^{i \phi}$ is not fixed in this case, leading to the $u(\psi)$ landscape sketched in \fig{ls} (left panel, "mexican-hat potential"). However, any nonzero value for $\vert \tilde{\gamma} \vert$ locks the phase of $\psi_0$ to the imaginary axis (i.e., $\phi$ = $\pm \pi/2$ and therefore $\psi_0$ = const.), and leads to an optimum value $n_c(0) = (\mu+2\vert \tilde{\gamma}\vert)/v_0$ in a "Napoleon's hat potential" (see \fig{ls}, right panel). The minimum potential energy becomes $u_{min} = -(\mu+2\vert \tilde{\gamma}\vert)^2/2v_0$ which is smaller than in the axially symmteric case. The saddle-point value for $u$ on the real axis for $\mu > 2\vert \tilde{\gamma}\vert$ is $-(\mu-2\vert \tilde{\gamma}\vert)^2/2v_0$. Note that these energy densities are all expressed per \t{dimer}. The corresponding energies per triplon are $u/n_c(0)$.
\\ A vanishing $n_c(0)$ is realized when $(\mu+2\vert \tilde{\gamma}\vert)$ = 0. \t{As a consequence}, the
gap field $H_c$ that would be observed in an ideal system with axial symmetry is renormalized to a value $H_c^{exp}$ = $H_c- \Delta H_c$ (with $ \Delta H_c$~=~$2\vert \tilde{\gamma}\vert/g\mu_0\mu_B$) above which condensation occurs. Taking a reasonable value for $\vert \tilde{\gamma}\vert \approx$ 0.01 meV for $\mathrm{TlCuCl_3}$ \cite{kolezhuk, sirker} and $\vec{H} \parallel b$ with $g$ =2.06 \cite{kolezhuk}, we obtain a renormalization of the critical field due to $\vert \tilde{\gamma}\vert$ alone by $\mu_0\Delta H_c \approx$ 0.2 T for this particlular magnetic-field direction.
\\The \t{resulting} condensate \t{fraction}, $n_c(0) = \mu^{exp}/v_0$ where $\mu^{exp} = g\mu_B\mu_0(H-H_c^{exp})$, \t{is} in full agreement with Hartree-Fock calculations for spin dimer systems \cite{Nik}. 
\begin{figure}
\includegraphics[scale=0.23]{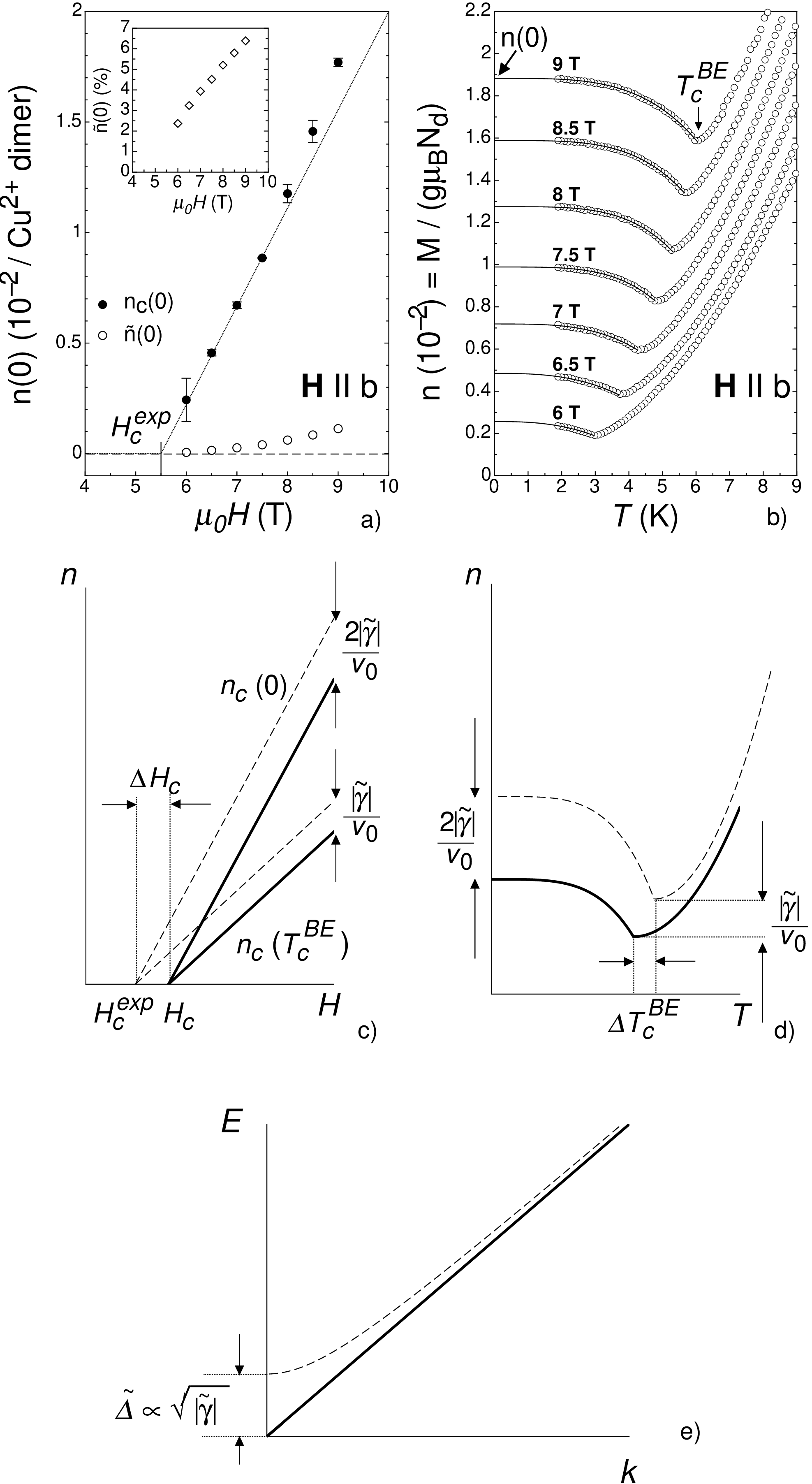}
\caption{\label{dens} a) Condensate \t{fraction} $n_c(0)$, non-condensed triplon \t{fraction} $\tilde{n}(0)$, and the percentage of non-condensed triplons (inset) \cite{we}. b) Triplon-\t{fraction} $n(0)$ data obtained from magnetization $M(T)$ data for $\vec{H} \parallel b$ (see arrow for the $\mu_0H$~=~9~T data and Ref. \cite{we}).  c) Schematic representations of the effects of violated axial symmetry on the triplon \t{fraction} at $T=0$ and at $T=T_c^{BE}$, d) on the triplon \t{fraction} at fixed magnetic field $H>H_c$ \cite{Nik, sirker, sirker2} and e) on the excitation spectrum $E(k)$ (axes are not to scale). Solid and dashed lines represent an axially symmetric system and a system with violated axial symmetry, respectively.} 
\end{figure}
In \fig{dens}a we show the triplon condensate \t{fraction} $n_c(0)$ as deduced from our magnetization $M(T,H)$-data of $\mathrm{TlCuCl_3}$ \cite{we}, see \fig{dens}b, that we have already corrected for a small \t{fraction} $\tilde{n}$ of noncondensed triplons \cite{Nik}. These data have been obtained from the simple relation $M(0)~=~g\mu_B n(0)N_d$ (with $N_d$ the number of dimers and $n(0)$ the total triplon \t{fraction} at $T$ = 0) \cite{Nik}, without assuming any specific value for $v_0$.
A linear fit to the data in the dilute limit ($\mu_0H_c < \mu_0H \lesssim$ 7.5 T \cite{sirker}) yields $\mu_0H_c^{exp} = 5.501\pm$ 0.003 T and $v_0/k_B$~=~311.4~$\pm$~0.5$\mathrm{Km^3}$ which is in very good agreement with available literature values \cite{sirker, misguich, yamada}. The deviation from the linear behavior at larger magnetic fields can be attributed to the contribution of higher triplet states \cite{mats}.
Our simple formalism does not include the influence of such triplet states, nor does it allow for a determination of $\tilde{n}$ itself, but this latter correction is of the order of a few percents at most in our data \cite{we, Nik}, as it is typical for a weakly interacting Bose gas, see inset of \fig{dens}a.
\begin{figure}
\includegraphics[scale=0.53]{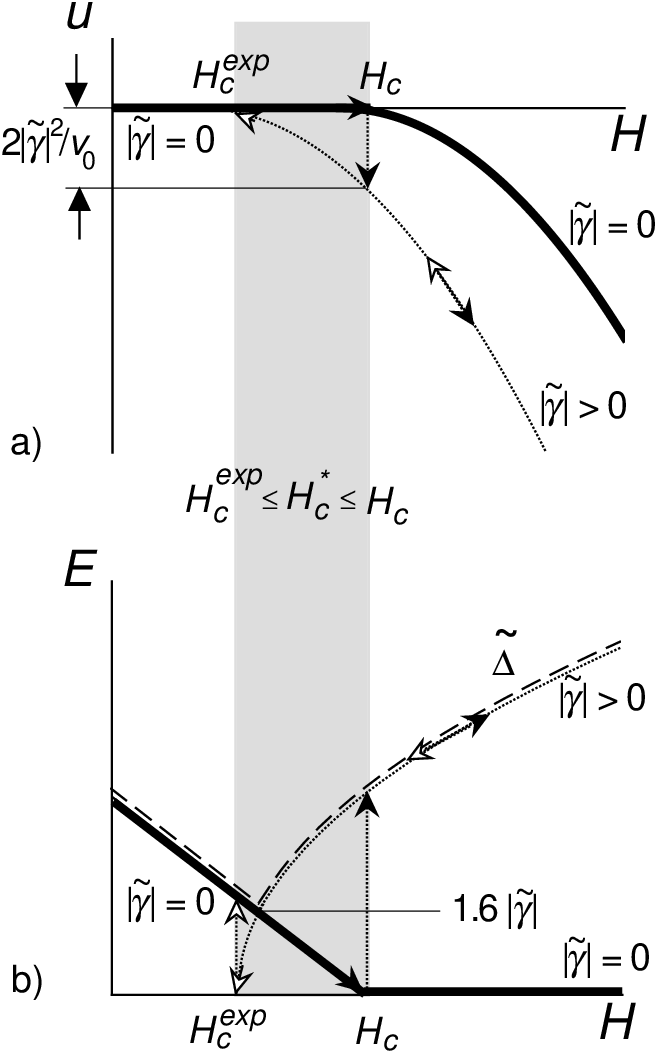}
\caption{\label{all} a) \tc{Potential energy per dimer}, \t{and b)} energy-level scheme of the lowest triplon branch of an axially symmetric system (\t{$\vert \tilde{\gamma} \vert =0$}, thick solid lines), and of a system that shows a spontaneous axial distortion (\t{$\vert \tilde{\gamma} \vert > 0$) above $H_c^\star$ (thin lines and arrows, axes are not to scale)}, respectively. \tc{The grey shaded area indicates the maximum hysteresis width of a possible weakly first-order transition at $H_c^\star$ with $H_c^{exp} \leq H_c^\star \leq H_c$ (filled arrows: $H$ increasing; open arrows: $H$ decreasing). The dashed line in b) reproduces the trend in the ESR data from Refs. \cite{kolezhuk} and \cite{glazkov} for $\mathrm{TlCuCl_3}$ and $\vec{H} \parallel b$ (see text).}}
\end{figure}
\\ \tt{Focusing further on} the effects \t{of an explicit violation of axial symmetry}, the condensate \t{fraction} at $T$ = 0 is changed by 2$\vert \tilde{\gamma}\vert/v_0$ in our calculation, see Figs. 2c and d. \tt{If we use} the fact that the total triplon fraction $n(0) \approx n_c(0)$ and \tt{take the result} $n(T_c^{BE})$ = $n(0)$/2 from Ref. \cite{Nik} \tc{assuming a quadratic triplon dispersion relation}, we obtain a shift $\vert \tilde{\gamma}\vert/v_0$ in $n(T_c^{BE})$ that is again in full agreement with the corresponding HF calculations \cite{sirker}, see again Figs. 2c and d.
\\We may relate the minimum value $u_{min}$ at $T$ = 0 to the transition temperature $T_c^{BE}$ if we assume that $\vert u_{min} \vert /n_c(0)$, the energy gain per triplon upon condensation, is proportional to $k_BT_c^{BE}$ with a field-dependent proportionality factor of the order of unity. Any nonzero $\vert \tilde{\gamma}\vert$ leads to an increase in $T_c^{BE}$ as compared to the axially symmetric case, see \fig{dens}d. This trend can be clearly seen in the calculated $M(T,H)$-curves from Ref. \cite{sirker}, where the minimum in $M$ that is usually taken as a criterion to define $T_c^{BE}$ is shifted towards higher temperatures \t{as soon as $\vert \tilde{\gamma}\vert > 0$}. If we again take $\vert \tilde{\gamma}\vert$ = 0.01 meV we obtain a shift in $T_c^{BE}$ of the order of $\Delta T_c^{BE} \approx \vert \tilde{\gamma}\vert / k_B \approx$ + 0.1 K, which has to be compared to the result of the more precise HF calculations with $\Delta T_c^{BE} \approx$ + 0.5 K  \cite{sirker}.
\\As a consequence of the anisotropy term $\vert \tilde{\gamma}\vert $ the original invariance of $u(\psi)$ with respect to a transformation $\psi \to \psi e^{i\phi}$ is lifted. The maximum variation in the potential-energy \t{per dimer} $u$ along the ellipsoid contour with local minima in the radial $\vert\psi\vert$-direction, see \fig{ls}, is $4\mu \vert \tilde{\gamma}\vert/v_0$, or $\mathit{\tilde{\Delta}}$~$\approx$ $4\vert \tilde{\gamma}\vert$ per ground-state triplon. This determines the order-of-magnitude of an anisotropy gap that can be calculated to $\mathit{\tilde{\Delta}}$~=~$\sqrt{8\vert\tilde{\gamma}\vert g\mu_B\mu_0(H - H_c)}$ \cite{sirker, us}. 
\\Excitations above the ground state $\psi_0$ with energies below $\mathit{\tilde{\Delta}}$ will clearly not show the typical gapless Goldstone-like behavior as expected for an axially symmetric system. For $\mathrm{TlCuCl_3}$ we calculate \t{with $\vert \tilde{\gamma} \vert=0.01$ meV} \cite{kolezhuk, sirker} a gap $\mathit{\tilde{\Delta}}$~$\approx$~0.3~meV for $\mu_0H$~=~14~T along $b$, which is somewhat below what has been resolved in inelastic-neutron scattering measurements \cite{rueg}. For excitation energies larger than $\mathit{\tilde{\Delta}}$ the presence of an anisotropy gap may remain unnoticed, see \fig{dens}e. 
\\ \tc{In $\mathrm{TlCuCl_3}$ such a gap may arise from a pre-existing anisotropy that is already present in $H$ = 0 \cite{glazkov}. In the following we argue, however, that even a perfectly axially symmetric magnetic system is unstable towards a spontaneous violation of this symmetry at $H_c$, which inevitably leads to the formation of a small anisotropy gap $\tilde{\mathit{\Delta}}$ above $H_c$ of real materials.}
\subsection{\label{susc}Instability of the condensate towards violation of axial symmetry}
A striking fact in our analysis is that the minimum potential energy per dimer is \textit{smaller} with a nonzero $\vert \tilde{\gamma}\vert$ than in an analogous axially symmetric system with $\vert \tilde{\gamma}\vert$ = 0. \tc{This means that a distortion of the original crystal symmetry may spontaneously occur at $H_c$ together with an increase of $\vert \tilde{\gamma} \vert$, provided that the total energy, including both magnetic and crystal-lattice contributions, is lowered along with this distortion. This argument is so general that it should be applicable to all insulating spin systems that are supposed to show a Bose-Einstein condensation of magnetic bosonic quasiparticles. As we do not not make any specific assumptions on the microscopic arrangement of the spin-carrying atoms, one cannot make any more precise, universal statement about the details of the resulting lattice distortion.} 
\\ \tc{The gain in potential energy per dimer upon condensation in combination with this simultaneous distortion is $2\vert \tilde{\gamma}\vert ^2 / v_0$, see \fig{all}a. If the critical field is approached from below with $H$ increasing, the parameter $\vert \tilde{\gamma}\vert$ may therefore jump discontinuously from zero to its optimum value either at $H_c$ or at a transition field $H_c^\star$ with $H_c^{exp} \leq H_c^\star \leq H_{c}$, while with $H$ decreasing the transition can take place at a different field in the same magnetic-field interval, see grey shaded area in Figs. \ref{all}a and b}. \t{In an ideal situation with a perfect axial symmetry in $H = 0$, the critical field $H_{c}$ corresponds to a "normal-state" value with $\vert \tilde{\gamma}\vert = 0$ that is determined only by the gap energy, while in the condensate phase the effective critical field is $H_c^{exp} = H_c - \Delta H_c < H_c$, lowered by $\Delta H_c = 2\vert \tilde{\gamma}\vert / g\mu_B\mu_0$ with respect to $H_c$ due to the increase in $\vert \tilde{\gamma}\vert$}. \tc{If the transition occurs at a transition field $H_c^\star$ that is strictly larger than $H_c^{exp}$, one will observe at $T=0$ corresponding small discontinuities in $u$ ($\Delta u \leq 2\vert \tilde{\gamma}\vert^2/v_0$, see \fig{all}a), $n_c$ ($\Delta n_c \leq 2\vert \tilde{\gamma}\vert/v_0$, see \fig{dens}c) and $M$ ($\Delta M = \Delta n_c g\mu_B N_d$), which would qualify the transition as of weakly first order with a maximum observable hysteresis width $\Delta H_c$. The occurrence of hysteretic effects in a real material may depend, however, on further conditions that are not considered here, such as material-quality issues or the relevance of possible quantum fluctuations at $T$ = 0 to the order of the phase transition.} 
\\ \tc{ Even if a non-zero $\vert \tilde{\gamma}\vert$ evolves continuously above the critical field as a function of $H$ along with a continuous structural distortion with no detectable hysteresis, one should still be able to distinguish between the Ònormal-stateÓ $H_c$ extracted from the experimental data taken below $H_c^{exp}$, and a $H_c^{exp} < H_c$ from corresponding data taken well above $H_c$, respectively, to obtain an estimate for $\vert \tilde{\gamma}\vert$ from the resulting difference $\Delta H_c$. In any case, if $H_c^{exp} < H_c^\star < H_c$, one expects a finite energy difference  $\leq 1.6 \vert \tilde{\gamma} \vert$  at the transition field $H_c^\star$ between the lowest-triplon state and the singlet states, respectively, see \fig{all}b.}
\\ \tc{A possible pre-existing anisotropy that may already be present at $H$ = 0 (which is likely to be the case in $\mathrm{TlCuCl_3}$) can be easily included in this formalism by identifying $H_c$ with a renormalized "normal-state" value that already contains this pre-existing anisotropy. Any additional $\vert \tilde{\gamma}\vert$ that may develop together with the lattice distortion around this critical field will somewhat change the value of the anisotropy gap $\tilde{\mathit{\Delta}}$. The difference ${\Delta H_c}$, however, and possible discontinuities in $u$, $n_c$ and $M$ are determined by the additional $\vert \tilde{\gamma}\vert$ alone.}
\\It is remarkable that electron-spin resonance (ESR) data taken on $\mathrm{TlCuCl_3}$ do indeed show a clearly gapped behavior at and above $\vec{H}=H_c^{exp} \parallel b$ \cite{kolezhuk, glazkov}, \t{as we sketched in \fig{all}b}. Moreover, the ESR frequencies due to the lowest triplon gap in the "normal" phase for $\vec{H} \parallel b$ extrapolate to zero at a somewhat larger $H_c$ (by $\mu_0H_c \approx 0.2$~T) than the square-root like gap that we attribute to $\mathit{\tilde{\Delta}(H)}$ in the condensate phase and that can be fitted nicely with a $\vert \tilde{\gamma} \vert~\approx$ 0.016 meV, see energy-level scheme in \fig{all}b. Our scenario may also be an explanation for the observed abrupt changes in the ${}^{35}$Cl quadrupole shift \cite{vyaselev} that has been interpreted as an indication of a weakly-first order lattice deformation, as well as for the pronounced hysteretic behavior (with a $\mu_0\Delta H_c \approx~0.2~-~0.3~$~T) of the observed peaks in the sound-attenuation data of $\mathrm{TlCuCl_3}$ at $H_c(T)$ \cite{sherman}. These observations may indicate that  $\vert \tilde{\gamma}\vert$ of $\mathrm{TlCuCl_3}$ is indeed larger for $H > H_c$ than well below this value.
\\The present picture may also account for the first-order like \tccc{features} that have been seen in the X-ray data of the spin-ladder compound $\mathrm{Cu_2(C_5H_{12}N_2)_2Cl_4}$ \cite{lori} and in the \tccc{magnetocaloric effect of the} axially symmetric $S=1$ system $\mathrm{NiCl_2-4SC(NH_2)_2}$ \cite{zapf} at the respective magnetic phase transitions. It is also not unreasonable to assume that the observed gap feature in the ESR-data of this latter compound at $\mu_0H$~=~8~T \cite{z} is also related to a possible lattice distortion. 
\\ Such a spontaneous distortion arising from the interplay \t{between the emerging} magnetic Bose-Einstein condensate \t{and} its host crystal \t{that lowers} the total energy, with a tendency to increase or \textit{even create} an anisotropy perpendicular to the external magnetic field above $H_c$ even in a perfectly axially symmetric system, \tc{is rather unique and is reminiscent of the spin-Peierls instability in one dimensional magnetic chains}. \tccc{A spin-Peierls-type scenario has indeed been suggested to explain the
NMR data of the spin-ladder compound $\mathrm{Cu_2(C_2H_{12}N_2)_2Cl_4}$ around its critical magnetic field} \cite{cale, maya}. \tccc{This instability} \t{should} be a universal feature of magnetic BEC systems at their magnetic phase transition, and it is not expected to occur in axially symmetric Bose gases \t{composed of real particles} such as superfluid $\mathrm{{}^4He}$ or atomic condensates, where the condensate cannot create an axial-symmetry breaking term by itself.
\\ In a microscopic picture, the \tccc{tendency} of a magnetic condensate \tccc{to spontaneously
violate} the axial symmetry \t{can be interpreted as} a natural consequence of the transverse magnetic ordering that develops in the condensate phase \t{and that locks to the crystal lattice} due to unavoidable magnetoelastic coupling \cite{Nik, tanaka, matsu}. \tc{As soon as the transverse magnetic moments point to a specific, energetically preferred crystal direction, the phase $\phi$ that is associated with the angle between these  moments and the crystal axes \cite{giaandrueg} is indeed fixed (in $\mathrm{TlCuCl_3}$ with an angle $\alpha \approx 39^\circ$ to the $a$-axis \cite{tanaka}), and the magnetic analogue to a supercurrent velocity $\mathbf{v}_s=\hbar/m^\star \nabla \phi$ (where $m^\star$ is the effective mass of a triplon) is naturally zero for excitation energies below the anisotropic gap $\tilde{\mathit{\Delta}}$.} This gap covers an excitation-frequency range that \t{is} crucial for experiments that rely on the existence of a long-lived phase coherent condensate \cite{snoke1}, such as the detection of a \tccc{long-lived} spin supercurrent \tcc{as in} $\mathrm{ {}^3He-B}$ \cite{spincurr, htan}, of macroscopic second-sound like oscillations as observed in superfluid $\mathrm{{}^4He}$ \cite{pesh} or of \tcc{stable} vortex-like structures as they have been observed \tcc{both in superfluid} $\mathrm{ {}^4He}$ \tcc{and} in atomic condensates \cite{vort}. To achieve a lifetime of the order of seconds for a phase coherent condensate, a corresponding anisotropy gap may not exceed a few femto electron volts.
\\
In the absence of a Goldstone mode \t{(for which all values of the phase $\phi$ have to be energeticaly equivalent)}, the quantity $\psi_0^\dag\psi_0$ in the zero-frequency limit does no longer represent a condensate \t{fraction}, but it is rather related to the order parameter characterizing the antiferromagnetic state \cite{Affl}.

\section{Conclusions}
We have analyzed the spontaneous symmetry breaking in a Bose gas of \t{magnetic bosonic quasiparticles in insulating spin systems} based on simple functional methods in the classical limit. Our results reproduce several results from earlier HF approximated computations \cite{Nik, sirker, sirker2}, and various experimental findings in $\mathrm{TlCuCl_3}$ such as the occurrence and the magnitude of an anisotropy gap \cite{kolezhuk, glazkov} and a weakly first-order like behavior at the magnetic phase transition \cite{vyaselev, sherman} can be explained. \tt{On the basis of an energetic argument} we expect that all magnetic BEC systems \t{in insulating spin systems} are intrinsically unstable towards a spontaneous anisotropic distortion perpendicular to the external magnetic field, \tc{which leads to the formation of an anisotropy gap that is seriously limiting the lifetime of a phase coherent condensate.}
\\ We thank to O. Sushkov, A. K. Kolezhuk and \tt{E. Ya. Sherman} for stimulating discussion. This work was partly supported by the Schweizerische Nationalfonds zur F\"orderung der wissenschaftlichen Forschung, Grant. No. 20-111653.








\bibliography{dellamorebiblio}

\begin{thebibliography}{42}
\bibitem[{DM()}]{DM}
\bibinfo{note}{{The inclusion of a DM-like anisotropy term
  $i\gamma(\psi-\psi^\dag)$ \cite{sirker} would result in strictly imaginary
  $\psi$ because $u(\psi)$ has to be real-valued. \t{A} Goldstone-\t{like} mode
  can therefore not be obtained within this formalism.}}
\bibitem[{us()}]{us}
\bibinfo{note}{{This expression can also be qualitatively obtained within our
  formalism if we take $\Delta u$~=~$4\mu\tilde{\gamma}/v_0$, the \t{maximum}
  energy difference along the ellipsoid contour sketched in \fig{ls}, to be
  approximately equal to $\mathit{\tilde{\Delta}}^2/v_0$. $1/ v_0$ corresponds
  to the \t{condensate}-fraction per energy $\partial n_c(0)/\partial \mu$ at
  the ground state $\psi_0$, and $\mathit{\tilde{\Delta}}$ represents a gap
  energy per condensed triplon that is needed to compensate for $\Delta u$ }}.
\end{thebibliography}

\end{document}